\documentclass[a4paper]{article}  
\usepackage{amsmath,amssymb} 
\usepackage{graphicx}

\def\journalfont{\rm}         
\def\jou#1{{\journalfont #1\ }}
\def\joudef#1#2{\def #1{\jou{\ignorespaces #2}}}

\joudef{\aaa}    { Astron.\ Astrophys.}
\joudef{\aip}    { Adv.\ Phys.}
\joudef{\adm}    { adv.\ math.}
\joudef{\aihpa}  { Ann.\ Inst.\ H.\ Poincar\'e A}
\joudef{\am}     { Ann.\ Math.}
\joudef{\apny}   { Ann.\ Phys.\ (N.Y.)}
\joudef{\arnps}  {Annu.\ Rev.\ Part.\ Sci.}
\joudef{\apj}    { Astrophys.\ J.}
\joudef{\cjp}    { Can.\ J.\ Phys.}
\joudef{\cmp}    { Commun.\ Math.\ Phys.}
\joudef{\cqg}    { Class.\ Quantum Grav.}
\joudef{\grg}    { Gen.\ Rel.\ Grav.}
\joudef{\gp}      {Geom.\ and Phys.}
\joudef{\ijmpd}  { Int.\ J.\ Mod.\ Phys.\ D}
\joudef{\ijtp}   { Int.\ J.\ Theor.\ Phys.}
\joudef{\invm}   { Invent.\ Math.}
\joudef{\jm}     { J.\ Math.}
\joudef{\jmaa}   { J.\ Math.\ Anal.\ Appl.}
\joudef{\jmp}    { J.\ Math.\ Phys.}
\joudef{\jpa}    { J.\ Phys.\ A}
\joudef{\mnras}  { Mon.\ Not.\ R.\ Ast.\ Soc.}
\joudef{\mpla}   { Mod.\ Phys.\ Lett.\ A} 
\joudef{\nature} { Nature}
\joudef{\nc}     { Nuovo Cim.}
\joudef{\ncb}    { Nuovo Cim. B}
\joudef{\npb}    { Nuc.\ Phys.\ B}
\joudef{\ph}     { Physica}
\joudef{\pla}    { Phys.\ Lett. A}
\joudef{\plb}    { Phys.\ Lett. B}
\joudef{\pr}     { Phys.\ Rev.}
\joudef{\prd}    { Phys.\ Rev.\ D}
\joudef{\prep}   { Phys.\ Rep.}
\joudef{\prl}    { Phys.\ Rev.\ Lett.}
\joudef{\prsla}  { Proc.\ Roy.\ Soc.\ Lond.\ A}
\joudef{\ptp}    { Prog.\ Theor.\ Phys.}
\joudef{\ptps}   { Prog.\ Theor.\ Phys.\ Suppl.}
\joudef\rmp      { Rev.\ Mod.\ Phys.}
\joudef\spj      { Sov.\ Phys.\ JETP}

\newcommand\thickbaselines{\baselineskip=20pt\lineskip=3pt\lineskiplimit=3pt}
\catcode`@=11
\def\cases#1{\left\{\,\vcenter{\thicknormalbaselines\m@th
             \ialign{$##\hfil$&\quad##\hfil\crcr#1\crcr}}\right.}
\def\matrix#1{\null\,\vcenter{\thickbaselines\m@th
    \ialign{\hfil$##$\hfil&&\quad\hfil$##$\hfil\crcr
      \mathstrut\crcr\noalign{\kern-\baselineskip}
      #1\crcr\mathstrut\crcr\noalign{\kern-\baselineskip}}}\,} 
\catcode`@=12   

\newcommand{\eprint}{\textsf} 

\newcommand\be{\begin{equation}} \newcommand\ee{\end{equation}} 
\newcommand\bd{\begin{displaymath}}\newcommand\ed{\end{displaymath}}

\newcommand\ts\textstyle

\def\undersim#1{\mathop{\vtop{\ialign{##\crcr
     $\hfil\displaystyle{#1}\hfil$\crcr\noalign
     {\kern1pt\nointerlineskip}\hbox{$\hfil\sim\hfil$}\crcr
     \noalign{\kern1pt}}}}}

\def\apriori{{\it a priori \/}} 
 \def\etal{{\it et al.\ }} \def\ie{{\it i.e.\ }}
\def\cf{{\it cf.\ }}

\newcommand{\ncd}{\newcommand} 
\ncd{\Dtwo}{\mathcal{D}}
\ncd{\Ttwo}{\mathcal{T}}
\ncd{\lagomdot}{{\mbox{\large$\cdot$}}}
\ncd{\lagomddot}{{\mbox{\large$\cdot\cdot$}}}
\ncd{\dotprime}{^{(\lagomdot}{}'{}^)}
\ncd{\ddr}[1]{\frac{{#1}'}{r'}}
\ncd{\calQ}{\mathcal{Q}}
\ncd{\calC}{\mathcal{C}}
\ncd{\calU}{\mathcal{U}}
\ncd{\calR}{\mathcal{R}}
\ncd{\hatu}{\hat{u}}
\ncd{\hatr}{\hat{r}}
\ncd{\nms}{\negmedspace} 
\ncd{\nts}{\negthickspace} 
\ncd{\mcl}[1]{\mathcal{#1}} 
\ncd{\beq} {\begin{equation}} 
\ncd{\eeq} {\end{equation}} 
\ncd{\BE} {\begin{eqnarray}} 
\ncd{\EE} {\end{eqnarray}} 
\ncd{\rarr} {\rightarrow} 
\ncd{\larr} {\leftarrow} 
\ncd{\uarr} {\uparrow} 
\ncd{\darr} {\downarrow} 
\ncd{\lrarr} {\leftrightarrow} 
\ncd{\lbeq}[1]  {\label{eq: #1}} 
\ncd{\refeq}[1] {(\ref{eq: #1})} 
\ncd{\mrm}    {\mathrm} 
\ncd{\nn}{\nonumber} 
\ncd{\mbf}[1] {{\mathbf #1}} 
\ncd\T{\frac{1}{2}h^{\mu\nu}p_\mu p_\nu} 
\ncd{\ms}{\mathstyle} 
\ncd{\ds}{\displaystyle} 
\ncd{\bmth}[1] {\mbox{\boldmath $#1$}} 
\ncd{\abs}[1] {|#1|} 
\ncd{\ubold}{\mathbf u}
\ncd{\Abold}{\mathbf A}
\ncd{\Bbold}{\mathbf B}
\ncd{\Mbold}{\mathbf M}
\ncd{\tsfrac}[2]{{\ts\frac{#1}{#2}}}
\ncd{\lagom}{\hspace{.6pt}}
\ncd{\muk}{k}
\ncd{\dumkonstant}{v_0}
\ncd{\tdelta}{\grave\delta}
\ncd{\tDelta}{\grave\Delta}
\ncd{\deltaH}{\delta_{\!H}}
\ncd{\txi}{\grave\xi}
\ncd{\Lie}{\mathcal{L}}
\ncd{\mrc}{$M$-$R$ curve}
\newtoks\reportnoregister \newtoks\eprintnoregister
\newcommand{\reportnumber}[1]{\reportnoregister={#1}}
\newcommand{\eprintnumber}[1]{\eprintnoregister={#1}}

\reportnumber{\mbox{}} 
\eprintnumber{\mbox{}} 

\newcommand{\reportid}{
   \begin{minipage}{17cm}\vspace{-7.2cm}
     \begin{flushright}
      {\normalsize \the\reportnoregister \\[-.2cm]
            \eprint{\the\eprintnoregister}}\vspace{0.2cm}
     \end{flushright}
   \end{minipage}\hspace{-17cm} }

\catcode`@=11   
\def\title#1{\gdef\@title{\reportid#1}}
\catcode`@=12   

\ncd{\gtnoll}{{}^{0\!}\eta}
\ncd{\gnoll}{{}^{0\!}\eta}
\ncd{\bgtnoll}{{}^{0\!}\pmb{\eta}}

\begin{document} 

\reportnumber{USITP 2003-08}
\eprintnumber{gr-qc/0309056}

\title{Elastic Stars in General Relativity: II. Radial perturbations}
\author{Max Karlovini\footnote{E-mail: \eprint{max@physto.se}}, Lars
  Samuelsson\footnote{E-mail: \eprint{larsam@physto.se}} \:and
  Moundheur Zarroug\footnote{E-mail: \eprint{moundheur@physto.se}}
  \\[10pt] {\small Department of Physics, Stockholm University} \\
  {\small AlbaNova University center, 106 91 Stockholm, Sweden} }
\date{} \maketitle


\begin{abstract}{\normalsize
    We study radial perturbations of general relativistic stars with
    elastic matter sources. We find that these perturbations are
    governed by a second order differential equation which, along with
    the boundary conditions, defines a Sturm-Liouville type problem
    that determines the eigenfrequencies. Although some complications
    arise compared to the perfect fluid case, leading us to consider a
    generalisation of the standard form of the Sturm-Liouville
    equation, the main results of Sturm-Liouville theory remain
    unaltered. As an important consequence we conclude that the
    mass-radius curve for a one-parameter sequence of regular
    equilibrium models belonging to some particular equation of state
    can be used in the same well-known way as in the perfect fluid
    case, at least if the energy density and the tangential pressure
    of the background solutions are continuous. In particular we find
    that the fundamental mode frequency has a zero for the maximum
    mass stars of the models with solid crusts considered in Paper I
    of this series.}
\end{abstract}
\vspace{.5cm}
\centerline{\bigskip\noindent PACS: 04.40.Dg, 97.10.Sj, 97.60.Jd}

\section{Introduction}

This is the second paper in a series that intends to bring the
relativistic theory of elastic matter sources to a point where it can
be conveniently used in numerical studies of compact stars. In the
first paper\cite{ks:relasticityI}, hereinafter referred to as Paper I,
we reviewed and extended the theory as well as applied it to static
spherically symmetric (SSS) configurations. As an illustration, we
constructed stellar models with elastic crusts or cores numerically
using a relativistic polytrope as the unsheared part of the equation
of state, while the shear modulus was assumed to be proportional to
the pressure. 

In this paper we study adiabatic radial perturbations of general
relativistic SSS stars with elastic matter sources. By an adiabatic
perturbation we effectively mean that the cold matter equation of
state that is used to model the eqillibrium configuration is assumed
to be valid also for the perturbed, dynamic, system. Thus we limit our
analysis to situations where the time scale of the perturbations is
much smaller than any time scale associated with reactions within the
medium. It should be mentioned here that this need not be a
good approximation for the matter in the deep interior of neutron
stars, for a discussion see \cite{ghg:maxmass} and references therein.

The study of stellar oscillations dates back to 1918 when Eddington
wrote down the equation for adiabatic radial oscillations of Newtonian
perfect fluid stars\cite{eddington:puls}. The first landmarks in the
study of the corresponding general relativistic case were the works of
Chandrasekhar\cite{chandra:dyninstgasletter,chandra:dyninstgas}
including his famous result on the dynamical instability of white
dwarfs\cite{chandra:dyninstwd}. Soon thereafter
Tooper\cite{tooper:apj65} wrote a very thorough paper on the
equilibrium and stabillity of stars obeying a relativistic polytropic
equation of state. One also has to mention the book by Harrison \etal\ 
\cite{HTWW:gravcollapse}. Since then, a huge pile of papers have been
written on the subject but we make no attempt to review the historical
progress of the field here. Some more recent references to previous
works can be found in the nice paper by Kokkotas and
Ruoff\cite{kr:radial} where a large set of frequencies is calculated
for realistic tabulated equations of state as well as classical
polytropes. Radial perturbations of SSS models with anisotropic
pressures have been studied before by Hillebrandt and
Steinmetz\cite{hs:stability} as well as Dev and
Gleiser\cite{dg:anisoII}. However, in those works \emph{ad hoc}
ans\"atze are made for how the matter responds to the perturbations.
Their results give qualitative insights into how pressure isotropy
affects the stability of SSS models, but when using a particular
matter description one wants to let that description govern the
equations. The Einstein equations for dynamical spherically symmetric
elastic models was notably studied by Magli in
\cite{magli:dynstructure}, but that work was not directly concerned
with small oscillations around static models.

This paper is organised as follows; In section \ref{sec:sss} we
consider radial perturbations of generic spherical matter sources
using a coordinate independent formalism. The only restriction we put
in at this time is that the stress-energy tensor should be
diagonalizable in an orthonormal frame. This is obviously a very mild
restriction, since most normal matter sources are believed to be of
this type. We take the somewhat non-standard approach of using the
Lagrangian (or comoving) gauge from the outset rather than starting
out from the Euler (or Schwarzschild) gauge only to transform to the
Lagrangian one along the way\cite{mtw:gravitation}. We then specialise
to a static background which of course simplifies the problem
considerably. Taking as our strategy to eliminate all occurrences of
the perturbation of the two-metric orthogonal to the SO(3) orbits we
end up with two equations relating the perturbations of the matter
degrees of freedom to the perturbation of the Schwarzschild radius.
These equations, being derived without specifying any particular
matter type, should prove useful for just about any situation where
radial perturbations of an SSS background are considered. Indeed,
every SSS metric corresponds to a stress-energy tensor which is
diagonal in an orthonormal frame and completely described by the
energy density, the radial pressure and the tangential pressure as
measured by a static observer. When radially perturbed, the
stress-energy tensor in fact always stays diagonalizable (to first
order) with respect to an orthonormal frame unless the energy density
and the radial pressure sum to zero, in which case the
diagonalizability may or may not continue to hold. For any physically
reasonable matter type that is used to model the interiors of stars,
this implies no restriction.

In section \ref{sec:boundary} we consider the boundary conditions
using the general Israel method of matching the first and second
fundamental forms. For the case of a static background on which the
energy density and the tangential pressure are continuous (the
background radial pressure must always be continuous) we find that
these conditions boil down to the matching of the Lagrangian
perturbations of the Schwarzschild radius and the radial pressure. If,
however, first order phase transitions in the energy density or
tangential pressure are present the matching is not uniquely
determined by the Israel junction conditions. Thus, in such
situations, one needs to make physical assumptions about the nature of
the phase transitions. 
At the surface of the star we find that the
Lagrangian perturbation of the radial pressure should vanish, whether
or not continuity of the energy density and the tangential pressure
hold there.

In section \ref{sec:relasticity} we finally specialise to elastic
matter using the formulation presented in Paper I. We find that, by
giving an elastic equation of state as an input, the problem reduces
to a single second order ordinary differential equation much in the
same way as for perfect fluids. Unlike the perfect fluid case,
however, the problem is not of the obvious Sturm-Liouville type, which
is closely related to the fact that the Lagrangian perturbation of the
radial pressure can not be written as a background function times the
first derivative of the perturbation function that for perfect fluids
is the unknown variable in the Sturm-Liouville equation.  The obvious
way of proving that one has a Sturm-Liouville problem at hand is to
make a transformation to the standard form of the Sturm-Liouville
equation. This can indeed be very easily done, but when doing so one
introduces the derivative of a background function which may have jump
discontinuites. For this reason we were not absolutely sure that our
perturbation equation with the imposed boundary conditions really
defines a system with all the standard Sturm-Liouville properties, so
we go through some lengths to prove that so is in fact the case under
quite general assumptions. Some parts of the proof concerning the
qualitative properties of the Pr\"ufer type phase function of the
system are somewhat technical and are therefore put in an appendix at
the end of the paper. We conclude this section with a discussion on
how the static approach to stability generalises to the case of
elastic matter.

Regarding the notation used in the present paper, we stick to the
notation of paper I as far as possible. In addition, we use $\Delta$
and $\delta$ to denote Lagrangian and Eulerian perturbation operators,
respectively. Whenever one of these operators appears in an equation,
it is to be understood that the equation is evaluated on the
background spacetime. 

\section{Radial perturbations of static spherically symmetric spacetimes}
\label{sec:sss}
Our approach to radial perturbations is to start out by writing the
full nonlinear Einstein equations for a general spherically symmetric
system in covariant two-dimensional language. To this end we denote
the metric on the SO(3) orbits by $t_{ab}$ ($t$ for tangential) and
the Lorentzian 2-metric on the orthogonal spaces $M_\bot$ by $j_{ab}$.
Furthermore, the curvature radius of $t_{ab}$ will be denoted by $r$
while the connection, Gaussian curvature and volume form of $j_{ab}$
will be referred to as $\Dtwo_a$, $K$ and $\epsilon_{ab}$,
respectively. We also introduce the mass function $m$ defined by
\begin{equation}\lbeq{Einmdef}
  1-\frac{2m}{r} = \Dtwo^a r\,\Dtwo_a r. 
\end{equation}
The four-dimensional Einstein tensor $G_{ab}$ can now be written as
(\cf appendix A of ref.\ \cite{pi:internal})
\begin{equation}
  r\,G_{ab} := 2\left[-\Dtwo_a \Dtwo_b r + \left(\Dtwo^c\Dtwo_c r -
  \frac{m}{r^2}\right)j_{ab} \right] + (\Dtwo^c\Dtwo_c r - rK)t_{ab}. 
\end{equation}
Thus the stress-energy tensor satisfying Einstein's equations $G_{ab}
= \kappa T_{ab}$ must be of the form
\begin{equation}
  T_{ab} = \Ttwo_{ab} + p_t\,t_{ab}, \quad \Ttwo_{ab}\,t^b{}_c = 0,
\end{equation}
where $p_t$ will be referred to as the tangential pressure. The
Einstein equations obviously split into 
\begin{align}\lbeq{eineqrad}
  -\Dtwo_a\Dtwo_b r + \left(\Dtwo^c\Dtwo_c r -\frac{m}{r^2}\right)j_{ab} &=
   \ts\frac12\kappa\,r\,\Ttwo_{ab} \\ \lbeq{pteq}
  \Dtwo^a\Dtwo_a r - rK &= \ts\kappa\,r p_t
\end{align}
Taking the trace of equation \refeq{eineqrad} and substituting the
result back we find
\begin{align}\lbeq{DDr}
  \Dtwo_a\Dtwo_b r &= \frac{m}{r^2}j_{ab} 
  - \tsfrac12\kappa r (\Ttwo_{ab}-\Ttwo j_{ab}) \\ \lbeq{EinK}
  K &= \frac{2m}{r^3} + \ts\frac12\kappa(\Ttwo - 2p_t)
\end{align}
It is worth noting that a contraction of eq.\ \refeq{DDr} with
$\Dtwo^b r$ gives the following expression for the gradient of 
the mass function:
\begin{equation}\lbeq{EinDm}
  \Dtwo_a m = \ts\frac12\kappa r^2(\Ttwo_a{}^b-\Ttwo j_a{}^b)\Dtwo_b r 
\end{equation}
The matter equations of motion $\nabla_b T^b{}_a = 0$ can be written in
two-dimensional language as 
\begin{equation}\lbeq{Einmatter}
  r^{-2}\Dtwo_b(r^2\Ttwo^b{}_a) - 2r^{-1}p_t\,\Dtwo_a r =
  \Dtwo_b\Ttwo^b{}_a + 2r^{-1}(\Ttwo_a{}^b - p_t j_a{}^b)\Dtwo_b r = 0.
\end{equation}
The fact that $K$ is supposed to be the Gaussian curvature associated
with the connection $\Dtwo_a$ can be expressed as
\begin{equation}\lbeq{DKconsistent}
  \epsilon^{ab}\Dtwo_a \Dtwo_b v_c = K\epsilon_c{}^a v_a,
\end{equation}
for any one-form $v_a$. Taking $v_a = \Dtwo_a r$ and using eq.\ 
\refeq{DDr} we may rewrite the left hand side of eq.\ 
\refeq{DKconsistent} as
\begin{equation}
  \epsilon^{ab}\Dtwo_a \Dtwo_b \Dtwo_c r =
  \left(\frac{2m}{r^3}+{\ts\frac12}\kappa\tau\right)\epsilon_c{}^a \Dtwo_a
  r - {\ts\frac12}\kappa r^{-1} \Dtwo_b(r^2\tau^b{}_a)\epsilon_c{}^a.
\end{equation}
Setting this expression equal to $K\epsilon_c{}^a\Dtwo_a r$ we find
that eq.\ \refeq{EinK} is in fact implied by eqs.\ \refeq{DDr} and
\refeq{Einmatter}. We shall use this fact to let the the latter two
equations provide the full Einstein equations for the spherically
symmetric system.

We shall now assume that $\Ttwo_{ab}$ is diagonalisable in an
orthonormal frame, viz
\begin{align}
  j_{ab} &= -u_a u_b + r_a r_b \\
  \Ttwo_{ab} &= \rho\,u_a u_b + p_r\,r_a r_b,
\end{align}
where $-u^a u_a = r^a r_a = 1$, $u^a r_a = 0$. We refer to this frame
as the comoving frame and the scalars $\rho$ and $p_r$ as the energy
density and the radial pressure, respectively. The symmetric
connection $\Dtwo_a$ is completely determined by the single connection
one-form
\begin{equation}
  \Gamma_a = r^b \Dtwo_a u_b = -u^b \Dtwo_a r_b. 
\end{equation}
In order to simplify the notation we introduce timelike and spacelike
derivatives according to
\begin{align}
  \dot{A} &:= u^a\Dtwo_a A \\
  A' &:= r^a\Dtwo_a A
\end{align}
for any scalar or tensor $A$. The connection one-form is in turn 
completely determined by the accelerations of the two frame vectors.
Indeed, it can be expressed as
\begin{equation}
  \Gamma_a = -(r^b\dot{u}_b)u_a - (u^br_b')r_a,
\end{equation}
where we note that the accelerations can be interpreted either as
two-dimensional or four-dimensional objects since
$\dot{u}_b=u^a\Dtwo_au_b=u^a\nabla_{\!a}u_b$ and
$r_b'=r^a\Dtwo_ar_b=r^a\nabla_{\!a}r_b$. Moreover we denote the frame
projections of the connection one-form by
\begin{align}
  \Gamma_0 &:= u^a\Gamma_a = r^a\dot{u}_a, \\
  \Gamma_1 &:= r^a\Gamma_a = -u^ar_a'.
\end{align}
With the dot-prime notation the definition of the mass function
\refeq{Einmdef} takes the form
\begin{equation}\lbeq{meq}
  -\dot{r}^2+r'^2 = 1-\frac{2m}{r}.
\end{equation}
Projecting eq.\ \refeq{DDr} with $u^au^b$, $r^ar^b$ and $u^{(a}r^{b)}$
we find
\begin{align}\lbeq{rdotdot}
  \ddot{r}-\Gamma_0 r' &= -\frac{m}{r^2}-\tsfrac12\kappa r p_r \\ \lbeq{rprimprim}
  r''-\Gamma_1\dot{r} &= \frac{m}{r^2}-\tsfrac12\kappa r\rho \\ \lbeq{rdotprim}
  r\dotprime - \tsfrac12 &\left(\Gamma_0\dot{r} + \Gamma_1 r'\right) = 0
\end{align}
where the symmetrization of the mixed derivatives of $r$,
$r\dotprime$, is needed because the derivatives do not commute. The
projections of the matter equations \refeq{Einmatter} take the forms,
\begin{align}\lbeq{rhodot}
  \dot\rho + (\rho+p_r)\Gamma_1 + 2(\rho+p_t)\frac{\dot{r}}{r} &= 0, \\ \lbeq{prprim}
  p_r' + (\rho+p_r)\Gamma_0 + 2(p_r-p_t)\frac{r'}{r} &= 0.
\end{align}
These last five scalar equations comprise the full Einstein equations
in spherical symmetry.  For completeness we also write down the
projections of the mass gradient equation \refeq{EinDm}. They are
simply
\begin{align}
  \dot{m} &= -\tsfrac12\kappa r^2 p_r \dot{r} \lbeq{mdot} \\
   m' &= \tsfrac12\kappa r^2\rho\,r' \lbeq{mprim}.  
\end{align}
If the spacetime is static it is easy to see that eqs.\ 
\refeq{rdotdot}, \refeq{rprimprim} and \refeq{prprim}, when expressed 
in Schwarzschild coordinates, reduce to the well-known Einstein
equations for a general SSS spacetime\cite{bl:aniso}, with eq.\
\refeq{prprim} then being the generalised TOV equation of hydrostatic
equilibrium when anisotropic stresses are allowed for. Furthermore
eqs.\ \refeq{rdotprim},
\refeq{rhodot} and \refeq{mdot} reduce to identities in the static
case.

\subsection{Perturbation formalism}

We shall apply a comoving, or \emph{Lagrangian} gauge, in which
$\Delta A$ can be interpreted as the perturbation of a quantity $A$ as
measured by an observer moving along with the four-velocity of the
matter. This is compatible with
\begin{align}
  \Delta u_a = \calU\,u_a, \\
  \Delta r_a = \calR\,r_a,
\end{align}
for some scalars $\calU$ and $\calR$. We take this to be the
definition of the gauge.\footnote{Strictly speaking this is not a
  \emph{proper} comoving gauge which would instead be defined by
  $\Delta u^a = 0$.} In contrast, one could also make the
Schwarzschild, or \emph{Eulerian}, gauge choice. Then $\delta A$
denotes the perturbation of $A$ as measured with respect to an
observer sitting at a fixed point in the Schwarzschild coordinate
grid. To discuss the connection between the two gauge choices it is
convenient to introduce the Schwarzschild frame given by
\begin{equation}
  j_{ab} = -\hatu_a\hatu_b + \hatr_a\hatr_b, 
\end{equation}
where
\begin{equation}
  \hatu^a\Dtwo_a r = 0, \quad -\hatu^a\hatu_a=\hatr^a\hatr_a=1, \quad 
  \hatu^a\hatr_a=0. 
\end{equation}
The relation between the comoving and the Schwarzschild frame is
uniquely specified by the boost velocity $v$ that relates the two
frames according to
\begin{align}
  \hatu^a &= \gamma(u^a-v r^a) \\
  \hatr^a &= \gamma(r^a-v u^a). 
\end{align}
where $\gamma=(1-v^2)^{-1/2}$.
The Schwarzschild gauge can be defined by
\begin{equation}
  \delta r = 0, \quad \delta\hatu_a = \hat\calU\,\hatu_a
\end{equation}
which implies $\delta\hatr_a = \hat\calR\,\hatr_a$, for some scalars
$\hat\calU$ and $\hat\calR$.  It is common practise to start out from
the Schwarzschild gauge and subsequently transform to the comoving
gauge.  We shall instead use the comoving gauge from the outset and
discuss the connection to the Schwarzschild gauge afterwards. Now, let
$\calC^c{}_{ab}$ be the background tensor which can be viewed as the
Lagrangian perturbation of the two-dimensional Christoffel symbols in
the sense that, for all one-forms $v_a$,
\begin{equation}
  \Delta(\Dtwo_{a}v_b) = \Dtwo_{a}(\Delta v_b) - v_c\calC^c{}_{ab}. 
\end{equation}
One can show that this tensor satisfies
\begin{align}
  r^br_c\calC^c{}_{ba} &= \Dtwo_a\calR \\
  u^bu_c\calC^c{}_{ba} &= -\Dtwo_a\calU,
\end{align}
from which it follows that
\begin{align}
  \Delta\Gamma_0 = \calU\,' - \Gamma_0\calR, \\
  \Delta\Gamma_1 = \dot{\calR} - \Gamma_1\calU.
\end{align}
Perturbing eq.\ \refeq{meq} we find
\begin{equation}
  \Delta\!\left(\frac{m}{r}\right) = \dot{r}\,\dot\xi 
  - r'\xi' - \dot{r}^2\calU + r'^2\calR,
\end{equation}
while perturbing eqs.\ \refeq{rdotdot}-\refeq{prprim} yields
\begin{align}
  \ddot\xi - \Gamma_0\xi' - \dot{r}\dot{\calU}-r'\calU\,' 
    - 2(\ddot{r}\calU-\Gamma_0 r'\calR) &=  
   -\frac1{r}\Delta\!\left(\frac{m}{r}\right) 
   + \frac{m}{r^3}\,\xi - \tsfrac12\kappa(r\Delta p_r+p_r\xi) \lbeq{xidotdot} \\
  \xi'' - \Gamma_1\dot\xi - \dot{r}\dot{\calR}-r'\calR' 
   - 2(r''\calR - \Gamma_1\dot{r}\calU) &= 
  \frac1{r}\Delta\!\left(\frac{m}{r}\right) 
   - \frac{m}{r^3}\,\xi - \tsfrac12\kappa(r\Delta\rho+\rho\,\xi)\lbeq{xiprimprim}\\
    \xi\dotprime -
    \tsfrac12 \left( \Gamma_0\dot\xi + \Gamma_1\xi' \right)
    - \dot{r}\,\calU\,'-r'\dot{\calR} &= 0,\lbeq{xiprimdot} 
\end{align}
\begin{align}
  (\Delta\rho)^\lagomdot + \Gamma_1(\Delta\rho+\Delta p_r) 
   + (\rho+p_r)\dot{\calR} + \frac{2\dot{r}}{r}(\Delta\rho+\Delta p_t) 
   + 2(\rho+p_t)\frac{r\dot\xi - \dot{r}\xi}{r^2} &= 0 \lbeq{deltarhodot}\\
  (\Delta p_r)' + \Gamma_0(\Delta\rho+\Delta p_r) + (\rho+p_r)\calU\,' 
   + \frac{2r'}{r}(\Delta p_r-\Delta p_t) 
   + 2(p_r-p_t)\frac{r\xi' - r'\xi}{r^2} &= 0 \lbeq{deltaprprim}
\end{align}
This is as far as we intend to proceed in full generality. In the next
subsection we specialise to the case of a static background with the
application to radially oscillating stars in mind. However it would be
interesting to use the above set of equations to study perturbations
of collapsing or exploding configurations. 


\subsection{Static background}

With a static background the equations given above simplify
considerably. The unperturbed four-velocity $u^a$ will then by
necessity (unless $\rho+p_r=0$) be aligned with the static Killing
vector $t^a$, \ie
\begin{equation}
  u^a = e^{-\nu}t^a, 
\end{equation}
where $\nu$ can be interpreted as the gravitational potential. This
means of course that $\dot{S} = 0$ for all unperturbed scalars $S$. It
also follows that $\Gamma_0$ is simply given by
\begin{align}
  \Gamma_0 = \nu', 
\end{align}
whereas $\Gamma_1$ vanishes identically since the acceleration $r_a'$
does. Consider first equation \refeq{xiprimdot} which simplifies to
\begin{equation}\lbeq{Rdot}
  \dot{\calR} = \frac1{r'}(\xi'^\lagomdot-\nu'\dot\xi).
\end{equation}
Hereinafter we shall set all static integration functions to zero
since they are only interesting when perturbing a static solution to a
static neighbour. Hence eq.\ \refeq{Rdot} integrates to
\begin{equation}\lbeq{B}
  \calR = \frac1{r'}(\xi'-\nu'\xi).
\end{equation}
Likewise, eq.\ \refeq{deltarhodot} is also a total dot derivative
whose integrated version is
\begin{equation}\lbeq{Deltarhofinal}
  \Delta\rho + \frac1{r'}(\rho+p_r)
  \left[\xi'-\left(\nu'-\frac{2r'}{r}\right)\xi\right] + \frac{6q}{r}\,\xi = 0, 
\end{equation}
where $q=\tsfrac13(p_t-p_r)$ and equation \refeq{Rdot} has been used.
We also solve eq.\ \refeq{xidotdot} for $\calU\,'$, with the result
\begin{equation}\lbeq{calUprim}
  \calU\,' = \frac1{r'}\left\{\tsfrac12\kappa r\Delta p_r + \ddot\xi + \nu'\xi' + 
  \left[\tsfrac12\kappa p_r - \nu'\left(2\nu'+\frac{r'}{r}\right) - 
  \frac{m}{r^3}\right]\xi \right\}
\end{equation}
Using eqs.\ \refeq{B}, \refeq{Deltarhofinal} as well as the
background equations we may rewrite \refeq{deltaprprim} according to
\begin{equation}\lbeq{Deltaprprimfinal}
\begin{split}
  -e^{-2\nu}r'\left(\frac1{r'}\,e^{2\nu}\Delta p_r\right)'
  +\frac{\rho+p_r}{r'}\left\{ -\ddot\xi +
    \left[ \nu'\left(\nu'+\frac{4r'}{r}\right)-\kappa p_r \right]\xi \right\}
  & \\ +\frac6{r}\left\{ q\left[\xi' +
      \left(\nu'-\frac{r'}{r}\right)\xi\right] + r'\Delta q \right\} &= 0. 
\end{split}
\end{equation}
These last four equations govern the dynamic radial perturbations of
any SSS background and regardless of the matter sources, provided that
the stress-energy tensor is diagonalizable in an orthonormal frame. As
already mentioned in the introduction, the condition of
diagonalizability is very unrestrictive since it is automatically
fulfilled if $\rho+p_r\neq 0$ holds on the background. For many matter
types the equation of state only imposes relations between the
perturbed matter quantities $\Delta\rho$, $\Delta p_r$, $\Delta q$ as
well as the perturbed Schwarzschild radius $\xi$. In such cases the
two equations \refeq{Deltarhofinal} and \refeq{Deltaprprimfinal}, from which all
reference to the perturbation of the two-metric $j_{ab}$ has been
eliminated, will give all the information needed. 

Let us now briefly discuss how to transform a perturbed quantity from
Lagrangian to Eulerian gauge, in case one would like to do such a
transformation. If the quantity is a tensor field the relation is
given by a vector field $X^a$ according to
\begin{equation}
  \delta = \Delta - \Lie_{X}.  
\end{equation} 
Since we are restricting ourselves to a static background the Lie
derivative will only be applied to static tensor fields. In fact we
only need to apply it to static scalars which means that it is
sufficient to determine the component $X^1=r_aX^a$ which according to
the gauge condition $\delta r=0$ is simply found from
\begin{equation} 
  0=\delta r= \xi -\Lie_X r=\xi-X^1r'.
\end{equation} 
So, for practical purposes we may set $X^a=(\xi/r')\,r^a$.  Consequently
the perturbation operators, when acting on scalars, are related
according to the well-known formula
\begin{equation} 
  \delta = \Delta - \xi\,\frac{\partial}{\partial r},
\end{equation} 
if Schwarzschild coordinates are used.

\section{Boundary conditions}\label{sec:boundary}
Quite generally, we want to consider the junction conditions across a
hypersurface $H$ of the perturbed spacetime which is given by
\begin{equation}
  F = \mathrm{constant},
\end{equation}
where $F$ is a spherically symmetric function. We assume that the
surface is timelike so that it has a spacelike unit normal which can
be written as
\begin{equation}
  n_a = \gamma_H(r_a - v_H\,u_a), \quad \gamma_H=(1-v_H^{\;2})^{-1/2},
\end{equation}
where $v_H=\dot{F}/F'$ can be interpreted as a boost velocity.  This
boost relates the two orthonormal frames $\{u^a, r^a\}$ and $\{w^a,
n^a\}$, where $w^a=\gamma_H(u^a - v_H r^a)$. The Israel junction
conditions imply that the first and second fundamental forms, $q_{ab}
= g_{ab} - n_a n_b$ and $K_{ab} = q_a{}^c q_b{}^d\nabla_{\!c}n_d$,
should be matched continuously across $H$. These quantities may be
expressed as
\begin{align}
  q_{ab} &= -w_aw_b + t_{ab} \\ 
  K_{ab} &= (w^cw^d\Dtwo_{\!c}n_d)w_aw_b
  - (r^{-1}n^c\Dtwo_c r)t_{ab},
\end{align}
where we have used the relation
\begin{equation}
 \nabla_a n_b = \Dtwo_a n_b - (r^{-1}n^c\Dtwo_c r)t_{ab}.
\end{equation}
This directly implies continuity of $w_a$ as well as of the scalar
quantities $r$, $n^a\Dtwo_a r$ and $w^cw^d\Dtwo_{\!c}n_d$. It will
prove convenient to replace the last of these quantities by
\begin{equation}
  J = G_{ab}n^an^b + \tsfrac12 R_H - r^{-2}, 
\end{equation}
where $R_H$ is the Ricci scalar of $H$. The continuity of $J$ follows
from the relation (\cf \cite{he:lss})
\begin{equation}
  R_H = -2G_{ab}n^an^b + (K^a{}_a)^2 - K^{ab}K_{ab}.   
\end{equation}
Using now that $G_{ab}n^an^b$ and $R_H$ can be expressed as
\begin{align}
  G_{ab}n^an^b &= \kappa\gamma_H^{\;2}(p_r+v_H^{\;2}\,\rho) \\
  R_H &= 4r^{-1}w^a\Dtwo_a(w^b\Dtwo_br) + 2r^{-2}[(w^a\Dtwo_ar)^2+1], 
\end{align}
we find that the Lagrangian perturbations of $w_a$, $r$, $n^a\Dtwo_a
r$ and $J$ are
\begin{align}\lbeq{P1}
  &\Delta w_a = \mathcal{U}\,u_a + \frac{(\Delta F)^\lagomdot}{F'}\,r_a \\
  &\Delta r = \xi \\ \lbeq{P2}
  &\Delta (n^a\Dtwo_a r) = \xi' - r'\calR = \nu'\xi \\ \lbeq{P3}
  &\Delta J 
  = \kappa\Delta p_r + 2r^{-1}\left(\xi-r'\frac{\Delta F}{F'}\right)^{\!\lagomddot}, 
\end{align}
where we have directly specialised to the case of a static background,
which for practical purposes implies that $H$ to zeroth order
coincides with a surface $r=\mathrm{constant}$ and that $v_H$ vanishes
when unperturbed. To obtain perturbed scalars that are to be
matched continuously over $H$ we make a gauge transformation to a
perturbation operator $\deltaH$ for which $\deltaH F = 0$, thus
corresponding to an observer that stays on $H$. This gauge
transformation is given by
\begin{equation}
  \deltaH = \Delta - \Lie_X, \quad X^a = X^0u^a + X^1r^a,\quad X^1 = \ds\frac{\Delta F}{F'},
\end{equation}
where $X^0$ has no \apriori restriction. Since the unperturbed values
of $w_a$, $r$, $n^a\Dtwo_a r$ and $J$ are $u_a$, $r$, $r'$ and $\kappa
p_r$, it now follows that
\begin{align}
  &\deltaH w_a = 
   \left[\mathcal{U}-\nu'\frac{\Delta F}{F'}-(X^0)^\lagomdot\right]u_a 
   + \left[\frac{(\Delta F)^\lagomdot}{F'} + (X^0)'\right]r_a \\
  &\deltaH r = \xi-\frac{\Delta F}{F'}\,r' \\
  &\deltaH (n^a\Dtwo_a r) = \nu'\xi - \frac{\Delta F}{F'}\,r''
  = \nu'\deltaH r + \tsfrac12\kappa r(\rho+p_r)\,\frac{\Delta F}{F'} \\
  &\deltaH J = \kappa\Delta p_r + 2r^{-1}(\deltaH r)^\lagomddot -
  \frac{\Delta F}{F'}\,\kappa p_r' = \kappa\left(\Delta p_r -
    \frac{6q}{r}\,r'\frac{\Delta F}{F'}\right) +
  \frac2{r}\left[\nu'\deltaH(n^a\Dtwo_a r) - (\nu')^2\deltaH r +
    (\deltaH r)^\lagomddot\right]
\end{align}
where background equations have been used. Regardless of the function
$F$, we can clearly make $r^a\deltaH w_a$ continuous by choosing the
function $X^0$ appropriately. We can also make $u^a\deltaH w_a$
continuous by making an appropriate matching of $\calU$ across the
boundary, which is possible since the perturbation equations only
determines $\calU'$ by eq.\ \refeq{calUprim}. Thus the continuity of
$\deltaH w_a$ can be assumed to have been taken care of. We now note
that the background quantities $r$, $r'$, $p_r$, $\nu$ and $\nu'$ are
required to be continuous, where the continuity of $p_r$ and $\nu'$
are equivalent. Moreover, when assuming a time dependence of the type
$e^{\pm i\omega t}$ for all perturbed scalars $S$, we can also use
that continuity for $S$ implies continuity for dot derivatives of $S$
(unless $\omega = 0$).  It thus follows that the junction conditions
for the radially perturbed solution imply continuity for the three
quantities
\begin{align}\lbeq{S1}
  S_1 &= \xi - \mathcal{F} \\ \lbeq{S2}
  S_2 &= (\rho+p_r)\mathcal{F} \\ \lbeq{S3}
  S_3 &= \Delta p_r - \frac{6q}{r}\mathcal{F},
\end{align}
where 
\begin{equation}
  \mathcal{F} = \frac{\Delta F}{F'/r'}. 
\end{equation}
These junction conditions in general depend on the choice of the
function $F$ that defines the boundary hypersurface $H$. Some
important special cases are worth mentioning here. If $\rho$ is
continuous on the background it follows from the continuity of $S_2$
that $\mathcal{F}$ is continuous which in turn implies that $\xi$ is.
If $\rho$ as well as the pressure anisotropy scalar $q$ are continuous
on the background, then so should both $\xi$ and $\Delta p_r$ be.
Moreover $\Delta p_r$ is always continuous over a phase boundary if
$q$ vanishes on both side of that boundary. This is of course always
the case for perfect fluids. 

For perfect fluids it interesting to consider the situation when the
background solution is such that $\rho$ has a jump discontinuity
occuring at some pressure $p_1$. In this case the Israel junction
conditions do not determine the quantity $\mathcal{F}$ across the
boundary. Therefore it must instead be determined by the
microphysics that governs the state of matter.  Then there are two
especially simple cases where the oscillation modes can be computed
without a detailed knowledge of that microphysics. 

The first alternative is to assume that the reaction rate is much
faster than the timescale associated with the perturbation. Then
matter can always be considered as being in thermodynamic equilibrium
which mean that the phase transition can be viewed as occuring at
$p_1$ in the perturbed star also. One should thence take the function
$F$ to be the pressure $p$ ($=p_r$), which gives
\begin{equation}
  \mathcal{F} = \frac{\Delta p}{p'/r'} 
  = -\frac{r(r-2m)\Delta p}{(\rho+p)\left(m+\tsfrac12\kappa r^2 p\right)}.
\end{equation}
Inserting this into eqs.\ \refeq{S1}-\refeq{S3} leads to continuity of
two quantities only, namely $\xi - (p'/r')^{-1}\Delta p$ and $\Delta
p$. In this case particles that are close to the phase boundary are
allowed to cross that boundary, meaning that observers that are
comoving with these particles will experience that the energy density
changes discontinuously.
As a consistency check of these boundary conditions, we have
numerically studied radial oscillations of models with a one-parameter
family of perfect fluid equations of state $\rho=\rho(p,\alpha)$ for
which $\rho$ develops a jump discontinuity as a function of $p$ when
the parameter $\alpha$ tends to a certain value $\alpha_1$. We found
that whereas $\xi$ also develops a jump discontinuity as
$\alpha\rarr\alpha_1$, neither $\xi-(p'/r')^{-1}\Delta p$ nor $\Delta
p$ do. 

The second alternative is to set $\Delta F = 0$, which implies
$\mathcal{F}=0$. This has the interpretation that fluid particles that
are sitting just inside or outside the boundary surface when the star
is unperturbed, will do so also when the star is oscillating. Hence no
particles jump between the two phases, which in turn means that one is
assuming that the transition reaction rate is much slower than the
timescale associated with the perturbation. Since $\mathcal{F} = 0$
implies continuity of $\xi$, the Israel junction conditions are in
this case satisfied simply by taking both $\xi$ and $\Delta p$
continuous. 

These two alternatives may be called \emph{fast} and \emph{slow phase
transitions}, respectively \cite{andrade:wmodes}. It is worth noting
that the slow phase transitions are straightforwardly generalised to
any type of matter; setting $\Delta F=0$ implies $\mathcal F=0$ which
means that the junction conditions are satisfied by requiring $\xi$
and $\Delta p_r$ to be continuous. However, there is no obvious way to
generalise the fast phase transitions to anisotropic matter.

At the surface of the star where we wish to match our oscillating
stellar model to the vacuum Schwarzschild solution with the same mass
$M$ as for the background solution, we set $\Delta F = 0$ since the
material particles at the surface are of course staying on the surface
during the oscillation. Thus $\mathcal{F} = 0$ and it follows that the
physically relevant boundary condition is
\begin{equation}\lbeq{surfacecond}
  \Delta p_r = 0. 
\end{equation}
It also follows that $\xi$ should be continuous, but since the frame
vectors $u^a$ and $r^a$ that are used to define $\Delta$ are not
uniquely defined in the exterior vacuum region, this implies no
further restriction at all. To make sure that we really get an
oscillating solution of the same mass as the static solution, we
should check that $\Delta m = 0$ at the surface. This will
automatically be the case since
\begin{equation}
  \Delta m = -\tsfrac12\kappa r^2 p_r\,\xi, 
\end{equation}
which vanishes when $p_r = 0$. 

To our knowledge the general matching conditions \refeq{S1}-\refeq{S3}
are new, and also the two above specialisations to perfect fluids
seems not to be well known.

At the centre we demand that the perturbed spacetime should be
regular, which implies the vanishing of
\begin{equation}
  \Delta\left(\frac{m}{r}\right) = -\left(\frac{m}{r^2}+\tsfrac12\kappa rp_r\right)\xi,
\end{equation}
leading to the condition
\begin{equation}\lbeq{regcenter}
  \lim_{r\rarr 0}r\xi = 0. 
\end{equation}

\section{Elastic matter}\label{sec:relasticity}
As described in Paper I we shall assume that the elastic matter
equation of state is set up using a fixed metric as the sole structure
on the three-dimensional material space $X$. Generically, this metric
has to be spherically symmetric to be compatible with a spherically
symmetric spacetime $M$, whether dynamic or static. Thus we may write
this metric as
\begin{equation}
  k_{AB}=\tilde{r}_A\tilde{r}_B+\tilde{t}_{AB},
\end{equation}
where $\tilde{t}_{AB}$ is the constant curvature two-metric on the
SO(3) orbits and $\tilde{r}_A$ is the outwards directed unit normal to
those orbits. Furthermore, compatibility with $M$ being spherically
symmetric also requires that the material space mapping $\psi$ should
give an isometry between the unit curvature metrics on the SO(3)
orbits of $M$ and $X$, which in turn implies that the pullback
$\tilde{t}_{ab} = \psi^*\tilde{t}_{AB}$ is
\begin{equation}
  \tilde{t}_{ab} = (\tilde{r}/r)^2\,t_{ab},  
\end{equation}
where $\tilde{r}$ is the curvature radius of $\tilde{t}_{AB}$. Since
$k_{ab} = \psi^*k_{AB}$ should be orthogonal to $u^a$ the same clearly
must be true for $\tilde{r}_a = \psi^*\tilde{r}_A$, viz
$u^a\tilde{r}_a = 0$ and $\tilde{r}_a = (r^b\tilde{r}_b)r_a$. Ignoring
the unphysical case when $\tilde{r}$ is constant we may write
$\tilde{r}_a = e^{\tilde\lambda}\Dtwo_a\tilde{r}$ for some function
$\tilde\lambda$ of $\tilde{r}$. Thus it follows that $\dot{\tilde{r}} =
0$ and 
\begin{equation}
  k_{ab} = (e^{\tilde\lambda}\,\tilde{r}\,')^2r_a r_b + (\tilde{r}/r)^2t_{ab},
\end{equation}
from which we read off that the linear particle densities (the square
roots of the eigenvalues of $k^a{}_b$) are given by
\begin{align}
  n_r &= e^{\tilde\lambda}\,\tilde{r}\,' \\ \lbeq{nt}
  n_t &= \tilde{r}/r.
\end{align}
Since $n_t$ is doubly degenerate we are, just as in the static case
treated in Paper I, effectively dealing with a two-parameter equation
of state $\rho=\rho_{\mathrm{eff}}(n_r,n_t)$ which comes from some
full three-parameter equation of state $\rho=\rho(n_1,n_2,n_3)$ by
setting $n_1=n_r$, $n_2=n_3=n_t$. The goal is here to specify the
final perturbation equations \refeq{Deltarhofinal} and
\refeq{Deltaprprimfinal} that are valid for a static background but
with general matter to the case of elastic matter. Since the equations
are expressed in terms of the matter variables $\rho$, $p_r$ and $q$,
it is convenient to work temporarily with $n=n_rn_t^{\,2}$ and $n_t$
as the two independent equation of state variables. The equation of
state then gives $p_r$ and $q$ according to
\begin{align}
  p_r &= n\frac{\partial\rho}{\partial n} - \rho \\
  q &= \tsfrac16\,n_t\frac{\partial\rho}{\partial n_t}. 
\end{align}
This in turn implies that the perturbation of $\rho$ can be expressed
as
\begin{equation}\lbeq{tdeltarho2}
  \Delta\rho = (\rho+p_r)\frac{\Delta n}{n} + 6q\frac{\Delta n_t}{n_t}. 
\end{equation}
It follows from eq.\ \refeq{nt} that
\begin{equation}\lbeq{Deltantxi}
  \frac{\Delta n_t}{n_t} = -\frac{\xi}{r}.
\end{equation}
Substituting this form of $\Delta n_t/n_t$ into eq.\ 
\refeq{tdeltarho2} and comparing the resulting expression for
$\Delta\rho$ with that implied by eq.\ \refeq{Deltarhofinal}, we can
solve for $\Delta n/n$ with the result
\begin{equation}\lbeq{Deltanxi}
  \frac{\Delta n}{n} = -\frac1{r'}\left[\xi' - \left(\nu' 
  - \frac{2r'}{r}\right)\xi\right]. 
\end{equation}
At this point it is convenient to replace $\xi$ with the variable 
\begin{equation}
  \zeta = r^2e^{-\nu}\xi.
\end{equation}
The expression for $\Delta n/n$ then compactifies to
\begin{equation}
  \frac{\Delta n}{n} = -r^{-2}e^\nu\frac{\zeta'}{r'}.
\end{equation}
This, in turn, implies that
\begin{align}\lbeq{Deltapr2}
  \Delta p_r &= \frac{\partial p_r}{\partial n}\,\Delta n +
  \frac{\partial p_r}{\partial n_t}\,\Delta n_t 
  = -r^{-2}e^\nu\left[ \beta_r\frac{\zeta'}{r'} 
  - 6\!\left(q-n\frac{\partial q}{\partial n}\right)\frac{\zeta}{r} \right] \\ \lbeq{Deltaq2}
  \Delta q &= \frac{\partial q}{\partial n}\,\Delta n +
  \frac{\partial q}{\partial n_t}\,\Delta n_t = 
  -r^{-2}e^\nu\left(n\frac{\partial
    q}{\partial n}\frac{\zeta'}{r'} + n_t\frac{\partial
    q}{\partial n_t}\frac{\zeta}{r}\right),
\end{align}
where we have used the Maxwell type relation
\begin{equation}
  n_t\frac{\partial p_r}{\partial n_t} = 6\!\left(n\frac{\partial q}{\partial n}-q\right),
\end{equation}
as well as the fact that the radial compressibility modulus $\beta_r$
in the present variables is given by
\begin{equation}
  \beta_r = n\frac{\partial p_r}{\partial n}. 
\end{equation}
We are now in the position to obtain a closed wave equation for
$\zeta$ by substituting eqs.\ \refeq{Deltapr2} and \refeq{Deltaq2}
into eq.\ \refeq{Deltaprprimfinal}. To display the result in a useful
and compact way it is convenient to introduce the auxiliary variable
\begin{equation}
  \eta = -\frac1{r'}\,e^{2\nu}\Delta p_r = P\left(\frac{\zeta'}{r'} - Y\zeta\right),
\end{equation}
where
\begin{align}
  P &= \frac{\beta_r e^{3\nu}}{r^2r'} \\
  Y &= \frac6{r\beta_r}\left(q-n\frac{\partial q}{\partial n}\right). 
\end{align}
Upon substituting eqs.\ \refeq{Deltapr2} and \refeq{Deltaq2} into eq.\ 
\refeq{Deltaprprimfinal} we now find that the radial perturbations of
SSS elastic models are governed by the single wave equation
\begin{equation}\lbeq{SLgen}
  -e^{2\nu}W\ddot\zeta + \frac{\eta\,'}{r'} + Y\eta + Q\zeta = 0, 
\end{equation}
where $\eta$, $P$ and $Y$ are given above and
\begin{align}
  W &= \frac{P}{v_{r||}^{\;2}\,r'^2}\,e^{-2\nu}, \\
  Q &= Q_1 + Q_2, \\
  Q_1 &= \frac{P}{v_{r||}^{\;2}\,r'^2}\left[\nu'\left(\nu'+\frac{4r'}{r}\right)
    -\kappa p_r\right], \\
  Q_2 &= P\left\{ \frac6{\beta_rrr'} \left[q\left(2\nu'-\frac{3r'}{r}\right) - n_t\frac{\partial q}{\partial n_t}\frac{r'}{r} \right] + Y^2 \right\},
\end{align}
with $v_{r||}^{\;2}=\beta_r/(\rho+p_r)$ being the squared speed of
radially directed longitudinal (sound) waves. While it was
convenient to use the variables $(n,n_t)$ to derive eq.\ 
\refeq{SLgen}, we shall now forget about them and change back to the
variables $(n_r,n_t)$ which makes $Y$ and $Q_2$ take the slightly
different forms
\begin{align}
  Y &= \frac6{r\beta_r}\left(q-n_r\frac{\partial q}{\partial n_r}\right) \\
  Q_2 &= P\left\{ \frac6{\beta_rrr'} \left[q\left(2\nu'-\frac{3r'}{r}\right) + 
  \left( 2n_r\frac{\partial q}{\partial n_r} - n_t\frac{\partial q}{\partial n_t} 
  \right)\frac{r'}{r} \right] + Y^2 \right\}
\end{align}
We also give the expressions for these quantities when the variables
$(n,z)$ are used, where $z$ is the shear variable $z=n_r/n_t$:
\begin{align}
  Y &= \frac6{r\beta_r}\left(q-n\frac{\partial q}{\partial n}-
  z\frac{\partial q}{\partial z} \right) \\
  Q_2 &= P\left\{ \frac6{\beta_rrr'} \left[q\left(2\nu'-\frac{3r'}{r}\right) 
  + 3z\frac{\partial q}{\partial z}\,\frac{r'}{r} \right] + Y^2 \right\}.
\end{align}
It is now easy to calculate $Y$ and $Q_2$ when using the particular
quasi-Hookean equation of state used in paper I. The result can be
displayed as
\begin{align}
  Y &= \frac4{r\beta_r}\left[\,\check\mu + 3\sigma + \tsfrac32(1-\check\Omega)q\,\right] \\
  Q_2 &= P\left\{ \frac6{\beta_rrr'} \left[q\left(2\nu'-\frac{3r'}{r}\right) 
  -2(\check\mu+3\sigma)\frac{r'}{r} \right] + Y^2 \right\},
\end{align}
where $\check\mu$ is the shear modulus, $\sigma$ the shearing energy
density and $\check\Omega = d(\ln\check\mu)/d(\ln n)$, as described in
paper I. Let us now abandon the covariant formalism and specialise to
Schwarzschild coordinates $(t,r)$ for the background. This is easily
done, since we shall only need that the dot and prime derivatives of
any scalar $S$ are given by
\begin{equation}
  \dot{S} = e^{-\nu}\frac{\partial S}{\partial t}, \quad S' = r'\frac{\partial S}{\partial r}, \end{equation}
with
\begin{equation}
  r' = e^{-\lambda} = \sqrt{1-\frac{2m}{r}}. 
\end{equation}
We also separate out the time dependence according to
\begin{equation}
 \zeta\rarr e^{\pm i\omega t}\zeta(r), \quad \eta\rarr e^{\pm i\omega t}\eta(r),
\end{equation}
which leads to the ordinary differential equation 
\begin{equation}\lbeq{SLgenSchw}
  \frac{d\eta}{dr} + Y\eta + (Q+\omega^2W)\zeta = 0, 
   \qquad \eta = P\left(\frac{d\zeta}{dr}-Y\zeta\right). 
\end{equation}
Clearly, this equation can easily be recast into the standard
Sturm-Liouville form
\begin{equation}\lbeq{SLstand}
  \frac{d}{dr}\left(P\frac{d\zeta}{dr}\right) + (\tilde{Q} + \omega^2W)\zeta = 0,
  \qquad   \tilde{Q} = Q - PY^2 - \frac{d(PY)}{dr}.
\end{equation}
Comparing the definitions of the coefficient functions of the two
equations \refeq{SLgenSchw} and \refeq{SLstand} with those of the
standard Sturm-Liouville equation that governs radial perturbations of
SSS perfect fluids (\cf \cite{mtw:gravitation}), we see that the
functions $P$, $W$ and the $Q_1$ part of $Q$ are direct
generalisations of the corresponding perfect fluid functions, while
$Y$ and the $Q_2$ part of $Q$ vanish identically when specialised to
the perfect fluid case. Since eq.\ \refeq{SLstand} is in
Sturm-Liouville form whereas \refeq{SLgenSchw} is not, one could think
that the former is the most useful way of writing the equation.
However, we shall now argue that this is not the case. We saw in
section \ref{sec:boundary} that $\xi$ and $\Delta p_r$ are contiuous
variables if $\rho$ and $q$ are continuous for the background solution
or, indeed, if we are considering slow phase transitions only. In that
case the variables $\zeta$ and $\eta$ will by their definitions be
continuous too. On the other hand, continuity of $\rho$ and $q$ does
not imply that other matter quantities occuring in the coefficient
functions of eqs.\ \refeq{SLgenSchw} and \refeq{SLstand} are
continuous. For instance, for the crust models considered in paper I
both $\rho$ and $q$ are continuous all the way to the surface of the
star but the shear modulus $\check\mu$ has a jump discontinuity at the
interface between the interior perfect fluid phase and the exterior
elastic one. This discontinuity in $\check\mu$ propagates to other
quantities such as the radial compressibility modulus $\beta_r$ and
hence the speed of sound $v_{r||}$ for radially propagating
longitudinal waves. In particular, for these models the jump
discontinuity of $\check\mu$ in this case implies a jump discontinuity
of $PY$, which in turn means that $Pd\zeta/dr = PY\zeta + \eta$ should
not be assumed continuous. Moreover $\eta$ has a more direct physical
interpretation than $Pd\zeta/dr$ as it is directly related to the
Lagrangian perturbation of the radial pressure and it is therefore
$\eta$ and not $Pd\zeta/dr$ that should vanish at the surface of the
star. For these reasons it is in general advantageous to work with
eq.\ \refeq{SLgenSchw} and to use $\zeta$ and $\eta$, rather than
$\zeta$ and $Pd\zeta/dr$ which is common when studying perfect fluid
stars\cite{kr:radial}, as the two independent variables when recasting
the second order differential equation \refeq{SLgenSchw} into a system
of two first order ones. As a bonus, it will not be necessary to
calculate the derivative $d(PY)/dr$ that enters the expression for
$\tilde{Q}$. The resulting first order system reads
\begin{align}\lbeq{dzetadr}
  \frac{d\zeta}{dr} &=  Y\zeta + P^{-1}\eta \\ \lbeq{detadr}
  \frac{d\eta}{dr} &= -\left[(Q+\omega^2 W)\zeta + Y\eta\right].
\end{align}
It should be supplemented by the boundary conditions at the centre $r=0$ and surface $r=R$
\begin{align}
  \lim_{r\rightarrow0}\frac{\zeta}{r}&=0 \\
  \lim_{r\rightarrow R}\eta &=0,
\end{align}
which follow from eqs.\ \refeq{regcenter} and \refeq{surfacecond}.

We now wish to determine how $\zeta$ and $\eta$ should behave near the
centre. To do so we again employ the quasi-Hookean class of equations
of state described in paper I. For this class it holds that
\begin{align}\lbeq{Pcenter}
  P &= P_0r^{-2} + O(r^0) \\
  Y &= Y_0r^{-1} + O(r) \\
  W &= W_0r^{-2} + O(r^0) \\ \lbeq{Qcenter}
  Q &= Q_0r^{-4} + O(r^{-2}),
\end{align}
where
\begin{align}
  P_0 &= e^{3\nu_c}\!\left(\check\beta_c+\tsfrac43\check\mu_c\right)>0,  \\
  Y_0 &= \frac{4\check\mu_c}{\check\beta_c+\tsfrac43\check\mu_c}, \qquad 0\le Y_0<3, \\
  W_0 &= e^{\nu_c}(\check\rho_c+\check{p}_c)>0, \\ \
  Q_0 &= -12e^{3\nu_c}
  \frac{\check\beta_c\check\mu_c}{\check\beta_c+\tsfrac43\check\mu_c}\le0       . 
\end{align}
The expansion coefficients $P_0$, $Y_0$ and $Q_0$ satisfy the relation
\begin{equation}\lbeq{constrel}
  Q_0 + P_0Y_0(3-Y_0) = 0, 
\end{equation}
which must be used to conclude from eq.\ \refeq{SLgen} that
$\zeta\sim\mathrm{constant}$ or $\zeta\sim r^3$ near the centre. The
regularity condition \refeq{regcenter} tells us that the latter
behaviour is the physical one. We hence conclude that the variables
$\zeta$ and $\eta$ leave the centre according to
\begin{align}\lbeq{zetac}
  \zeta &= Cr^3 + O(r^5)  \\ \lbeq{etac}
  \eta &= CP_0(3-Y_0) + O(r^2) \ldots,
\end{align}
where $C$ is an arbitrary constant. Although we have derived this
result for a restricted class of equations of states, one should
recall that the shear is zero at the centre of the star and thus small
in the vicinity of the centre. Hence the result should generalise to
all equations of state with a physically reasonable behaviour in the
limit of small shear, since the given class gives the most general
physically reasonable behaviour in that limit. 

Now, the first order system consisting of eqs.\ \refeq{dzetadr} and
\refeq{detadr} is in general not a standard Sturm-Liouville system due
to the presence of the function $Y$. Moreover the function $Q$ has a
non-standard $r^{-4}$ behaviour near the centre, which however goes
over into the standard $r^{-2}$ behaviour precisely when the leading
order $r^{-1}$ term of $Y$ vanishes. Still, it is not difficult to
show that the main Sturm-Liouville results for the standard case will
remain valid for our somewhat more general setting, assuming
continuity of $\zeta$ and $\eta$. To see that all eigenvalues
$\omega_n^{\,2}$ are real, let $(\zeta_n,\eta_n)$ and
$(\zeta_{n'},\eta_{n'})$ correspond to the eigenfrequences $\omega_n$
and $\omega_{n'}$, respectively.  Using eqs.\ \refeq{dzetadr} and
\refeq{detadr} one readily finds that 
\begin{equation}
  \frac{d}{dr}(\bar\zeta_n\eta_{n'}-\zeta_{n'}\bar\eta_n) = 
  (\bar\omega_n^{\,2}-\omega_{n'}^{\,2})\,W\bar\zeta_n\zeta_{n'}. 
\end{equation}
Integrating this equation over $0<r<R$ gives
\begin{equation}
  \left.(\bar\zeta_n\eta_{n'}-\zeta_{n'}\bar\eta_n)\!\right|_{\,r=0}^{\,r=R} = 
  (\bar\omega_n^{\,2}-\omega_{n'}^{\,2})\int_0^R\!W\bar\zeta_n\zeta_{n'}dr
\end{equation}
Since the boundary conditions always make the function
$\bar\zeta_n\eta_{n'}-\zeta_{n'}\bar\eta_n$ (the generalised
Wronskian) vanish at $r=0$ and $r=R$, it follows from setting $n'=n$
that $\bar\omega_n^{\,2}=\omega_n^{\,2}$ for all $n$, since $W>0$ for
$0<r<R$.  Since the eigenvalues thus indeed are real we can also
choose the eigenfunctions $\zeta_n$ to be real and conclude that
\begin{equation}
  \int_0^R\!W\zeta_n\zeta_{n'}dr = 0\quad 
  \mbox{for $\omega_n^{\,2}\neq\omega_{n'}^{\,2}$},
\end{equation}
\ie eigenfunctions corresponding to distinct eigenvalues are
orthogonal with respect to the interval $0<r<R$ and the weight
function $W$. To see that the eigenvalues are nondegenerate, form an
infinite discrete sequence which is bounded from below but not from
above, and can be enumerated such that
\begin{equation}
  \omega_0^{\,2} < \omega_1^{\,2} < \omega_2^{\,2} < \ldots, 
\end{equation}
with each corresponding $\zeta_n$ having exactly $n$ internal
($0<r<R$) separated nodes, it is convenient to make the Pr\"ufer type
substitution
\begin{equation}
  \zeta = \mathcal{A}\sin\varphi, \quad \eta = \mathcal{A}\cos\varphi.
\end{equation}
This leads to a first order differential equation for the phase
$\varphi$, reading
\begin{equation}
  \frac{d\varphi}{dr} = 
  P^{-1}\cos^2\!\varphi + Y\sin{2\varphi} + (Q+\omega^2 W)\sin^2\!\varphi. 
\end{equation}
Since $\tan\varphi = \zeta/\eta\rarr 0$ as $r\rarr 0$, we can declare
that $\varphi$ should leave the centre according to
\begin{equation}\lbeq{phicenter}
  \varphi = \frac1{P_0(3-Y_0)}\,r^3 + O(r^5), 
\end{equation}
which directly follows from eqs.\ \refeq{zetac} and \refeq{etac}.
Using the observation that $P$ is strictly positive on $0<r<R$, it is
immediate that
\begin{equation}
  \frac{d\varphi}{dr} = P^{-1} > 0, 
\end{equation}
at internal radii where $\varphi$ is an integer multiple of $\pi$, \ie
where $\zeta$ has internal nodes. This implies that the $\zeta=0$ line
of the $\eta$-$\zeta$ plane is always crossed in a counterclockwise
direction and that the nodes of $\zeta$ are separated since the origin
can never be encountered unless we are dealing with the trivial
solution $\zeta\equiv\eta\equiv 0$. Based on the fact that $W$ is also
strictly positive on $0<r<R$, it is fairly easy to show that the
phase, at each fixed internal radius $r$, is a continuous, monotonically
increasing function of $\omega^2$ such that
\begin{align}
  \lim_{\omega^2\rarr\infty}\varphi(r,\omega^2) = \infty \\
  \lim_{\omega^2\rarr-\infty}\varphi(r,\omega^2) = 0. 
\end{align}
This is exactly what is needed since it implies that the phase
function $\varphi_n$ corresponding to $\zeta_n$ will then satisfy
\begin{align}
  \varphi_n(r_{n'}) &= n'\pi, \quad n' = 1,2,\ldots,n \\ \lbeq{phaseR}
  \lim_{r\rarr R}\varphi_n(r) &= \left(n+\tsfrac12\right)\pi
\end{align}
where $r_{n'}$ is the radius of the $n'$:th node of $\zeta_n$. The limit
\refeq{phaseR} corresponds to the boundary condition at the surface of
the star being
\begin{equation}
  \lim_{r\rarr R}\eta = 0.
\end{equation}
Proofs of these statements on how $\varphi(r,\omega^2)$ depends on
$\omega^2$ are presented in the appendix. The proofs are based on
continuity for $\zeta$ and $\eta$, but it should be remarked
that they do not fail to be valid when the coefficient functions
$P$, $Y$, $Q$ and $W$ have a finite number of jump discontinuities.

The above results are very pleasing as they tell us that the radial
perturbations of elastic stars is in essence a simple Sturm-Liouville
problem that is a direct generalisation of the corresponding problem
for perfect fluid stars. 

To conclude this section it is interesting to note that, like the
perfect fluid case\cite{btm:catalogue}, the eigenvalues may be derived
from a variational principle. To this end we observe that the linear
system \refeq{dzetadr}~-~\refeq{detadr} follows from the
one-dimensional ``time''-dependent Hamiltonian
\begin{equation}
  H = \tsfrac12\left[P^{-1}\eta^2+2Y\zeta\eta+(Q+\omega^2W)\zeta^2\right].
\end{equation}
Indeed, since 
\begin{align}
  \frac{d\zeta}{dr} &= \frac{\partial H}{\partial\eta} \\
  \frac{d\eta}{dr} &= -\frac{\partial H}{\partial\zeta},
\end{align}
we see that $\eta$ can be interpreted as the conjugate momentum of
$\zeta$. From the corresponding Lagrangian, 
\begin{equation}
  L = \frac12\left[P\!\left(\frac{d\zeta}{dr}-Y\zeta\right)^{\!\!2}
    - (Q+\omega^2 W)\zeta^2\right]
\end{equation}
and standard methods\cite{arfken:math}, it can be readily shown that
the extremal values of the quantity
\begin{equation}
  \omega^2 = \frac{\displaystyle\int_0^R\left[P\!\left(\frac{d\zeta}{dr}-Y\zeta\right)^{\!\!2} - Q\zeta^2\right]dr}{\displaystyle\int_0^RW\zeta^2dr} 
\end{equation}
under variations of $\zeta$ that keep the boundary conditions
satisfied, are precisely the eigenvalues $\omega_n^{\,2}$.

\subsection{Stability and the mass-radius curve}

A technique which has proven very useful for stability considerations
of perfect fluid SSS stars is based on the analysis of static
configurations of a whole family of stars corresponding to a specified
equation of state. Given appropriate boundary conditions, for a given
central pressure $p_c$, there exist a unique solution to the TOV
equations, with a corresponding gravitational mass $M(p_c)$ and total
(Schwarzschild) radius $R(p_c)$. The set of equilibrium configurations
of a particular EOS can thus be represented as a curve in the $M$-$R$
plane. If shells of infinite density are disallowed the curve will be
continuous. Moreover, in the absence of first order phase transitions
the curve is guaranteed to be $C^1$\cite{lindblom:phase} (viewed as a
submanifold of the $M$-$R$ plane). Postponing for the moment the
discussion of possible implications of different kinds of phase
transitions we indicate how the \mrc\ can be used to determine the
stability of a given star. The original arguments were found by
J.~A.~Wheeler in the context of cold white dwarf and neutron star
models and are presented in \cite{HTWW:gravcollapse} (but see also
\cite{thorne:struct}).

As a first observation, note that since the radially perturbed models
are matched to the same exterior Schwarzschild solution as the
corresponding background models, it follows that static perturbations
($\omega_n^{\,2}=0$ for some $n$) are allowed only at extremum points
($dM/dR=0$) of the \mrc, since only at such points can static models
be perturbed to static neighbours of the same mass. Clearly then,
owing to the continuity of the eigenfrequencies viewed as functions of
the central pressure, the stability of the star can only change at
such extrema. Moreover, general arguments following from the
Sturm-Liouville properties of the perturbation equation show that the
mode changing stability at the extremum is even ($n=0,2,\ldots$) if
$dR/dp_c<0$ and odd ($n=1,3,\ldots$) if $dR/dp_c>0$. For equations of
state whose low pressure behaviour is known to give stable
configurations (this obviously includes neutron stars since small iron
balls are stable), it follows as a corollary that a mode is
(de)stabilised if the curve turns (counter)clockwise at an extremum.

Due to the general nature of the arguments they can be directly
applied also to the elastic case once the uniqueness and
Sturm-Liouville properties have been established. It should be clear
from the results of Paper I (see also
\cite{park:elastsss,bs:relasticity}) that, in the absence of jump
discontinuities in the energy density and/or the tangential pressure,
the equations of stellar structure possess a unique solution for each
central pressure. Although no proof is presented here it is fairly
obvious that the \mrc\ is $C^1$ under these restrictions. Moreover,
the Sturm-Liouville properties have been rigorously demonstrated in
this case so we conclude that the static stability analysis is
straightforwardly generalised to the elastic case.

However, in situations with first order phase transitions in the
density and/or the tangential pressure, the situation is more subtle.
For perfect fluids only the energy density can be be discontinuous
implying that the matching is uniquely determined by the junction
conditions. However, when there is a transition to an exterior elastic
phase, the tangential pressure just outside the phase boundary is not
specified by the junction conditions, as was shown explicitly in paper
I for the case of a perfect fluid interior to a quasi-Hookean type
elastic phase. This arbitrariness can be removed by considering the
value of the pressure anisotropy to be given by a parameter specified
as a part of the equation of state. If this view is taken, each
central pressure corresponds to a unique solution. In the presence of
such first order phase transitions the \mrc\ is well defined,
continuous, but may not be $C^1$. For the crust models presented in
Paper I with \mrc\ and compactness plots, the arbitrariness was
removed by simply setting the pressure anisotropy to zero on the outer
side of the phase boundary, which resulted in models with continuous
energy density and tangential pressure. Thus the static \mrc\ approach
to stability can definitely be used for these models.

Had we instead made other choices for the pressure anisotropy (in fact
we did play around with other choices, as mentioned in the paper) the
situation would be more unclear since 1) the \mrc\ may not be $C^1$
and 2) we have not proved that we are dealing with a Sturm-Liouville
type problem. The same is true for the models with elastic cores that
were also presented in Paper I, as well as for more general models for
which $\rho$ and $p_t$ are not continuous. 

For a perfect fluid with a jump discontinuity in the density at some
transition pressure $p_1$, the behaviour of the \mrc\ was studied by
Lindblom\cite{lindblom:phase}. Parametrising the discontinuity by
$\Delta=(\rho_+-\rho_-)/(\rho_-+p_1)$ where $\rho_\pm$ are the
densities just above and below the phase transition, he found that
there exists a critical value of this parameter
$\Delta_c=(\rho_-+3p_1)/2(\rho_-+p_1)$ where the slope of the \mrc\ 
has a discontinuity. Beyond the critical point the curve actually
turns $180^\circ$, leaving its tangent continuous. A similar analysis
as Lindblom's for elastic matter should be very useful, but is
postponed for later work.

Of course, if one is uncertain about the stability of a given model
one should always calculate the eigenvalues. For the relativistic
polytropic stellar models with quasi-Hookean elastic crusts
considered in Paper I we computed the first few modes and confirmed
that the fundamental mode did indeed have a zero at the maximum of the
\mrc. For the core models, we do not know theoretically that instability 
should set in at the maxima of the \mrc s, since the tangential
pressure and energy density are in fact discontinuous at the
boundary. However, assuming that the phase transition is slow, we find
that these models become unstable at, or at least very close to, the mass
maximum.

\section{Conclusions/discussion}\label{sec:conclude}

With the results of the present paper at hand we are now able to
numerically determine the frequencies of adiabatic radial
perturbations of general relativistic SSS stars with elastic matter
sources, without much further ado than in the much studied special
case of perfect fluid stars. For instance, both numerical schemes
employed by Kokkotas and Ruoff\cite{kr:radial} (based on the shooting
method and the Numerov method) are very easily modified to allow for
elasticity. Moreover, we established that the mass-radius curve, under
certain conditions, can be used to assess stability without having to
calculate a single frequency.

As expected, the spectrums of the radial perturbations of elastic SSS
models with moderate shear moduli do not differ much from the
corresponding spectrum when the shear modulus is identically vanishing,
as is the case for perfect fluid models.

Although a thorough treatment of general phase transitions is
postponed to later work, it seems clear that modest transitions do not
have a large impact on the stability properties. The connection
between large phase transitions and stability has, to our knowledge,
not been conclusively established even for perfect fluids. It would be
very useful to determine the precise relation between the behaviour of
the $M$-$R$ curve and the sign of the eigenfrequencies, for slow as well
as for fast phase transitions. We feel that it would probably be wise
to treat the perfect fluid case first, since the elastic case is
complicated by the additional freedom of specifying how the transition
is to be made.  It deserves to be mentioned that, as far as the crust
of a real neutron star is concerned, the density discontinuities are
not thought to have $\Delta$ larger than about
$10^{-1}\Delta_c$\cite{hp:expmatter}.  However, exotic phases in the
cores of neutron stars might lead to larger discontinuities. An
extreme example is provided by a strange quark star with a normal
matter envelope, in which case $\Delta$ can be as high as
$10^3\Delta_c$ or more\cite{glendenning:compactstars}.

Since we now have the spherically symmetric equilibrium configurations
as well as their stability under control, we are ready to take on more
complicated, and hence more interesting problems. The first step is of
course to treat non-radial first order perturbations of various kinds.
Work on time-dependent axial perturbations of prestressed
configurations is already on the way\cite{ks:relastaxial} applying the
general formalism of Karlovini\cite{karlovini:axial}. We also plan to
perturb SSS models into stationary neighbours, by setting them into
rotation as well as changing their shapes, the latter being impossible
for perfect fluid stars (to first order). What remains as far as first
order perturbations are concerned is the time-dependent polar sector.
However, unlike the situation in the axial sector, the matter and
metric perturbations (to the extent that they can be talked of
separately) are coupled in a highly non-trivial way already for
perfect fluids, which would probably make the effects of elasticity
very hard to assess. It therefore seems much more feasible as well as
interesting to instead directly treat rotation to second order in the
spirit of Hartle and Thorne\cite{hartle:slowrot, ht:slowrot}. Indeed,
such a setting would provide a sound basis for discussing very
interesting astrophysical phenomena such as free precession, quakes
and glitches of neutron stars, all of which are observed in
pulsars\cite{hp:nstheory}.

Although of less astrophysical interest, it would be fun to find some
exact SSS solution with a physically reasonable equation of state that
is found to be stable using the results of this paper.  We have in
fact recently found a few solutions with regular centres, finite
radii, everywhere causal wave propagation and positive energy
densities and eigenpressures\cite{ks:exact}. One of these solutions has the central
pressure as a free parameter and does consequently possess a $M$-$R$
curve. Using the formalism of the present paper we found that instability
sets in at the maximum of the $M$-$R$ curve as expected.

\appendix
\section*{Appendix}\label{sec:append}

In this appendix we prove that the phase function $\varphi$, satisfying the
differential equation
\begin{equation}\lbeq{phaseeq}
  \frac{d\varphi}{dr} = P^{-1}\cos^2\!\varphi + Y\sin{2\varphi} +
  (Q+\lambda W)\sin^2\!\varphi =: F(r,\varphi,\lambda), \quad \lambda
  = \omega^2,
\end{equation}
is a continuous, monotonically increasing function of $\lambda$ that
satisfies
\begin{align}\lbeq{lim1}
  \lim_{\lambda\rarr\infty}\varphi(r,\lambda) = \infty \\ \lbeq{lim2}
  \lim_{\lambda\rarr-\infty}\varphi(r,\lambda) = 0. 
\end{align}
Since eq.\ \refeq{phaseeq} is singular at $r=0$ and possibly also at
$r=R$, we must restrict ourselves to prove the statements for values
of $r$ in the open interval $0<r<R$ only.

The continuity directly follows from the continuity of
$F(r,\varphi,\lambda)$.  To prove monotonicity we observe that the
second term in the expansion \refeq{phicenter} for $\varphi$ is
\begin{equation}
  \frac{\lambda W_0}{5P_0^{\;2}(Y_0-3)^2}\,r^5. 
\end{equation}
Thus it follows that given $\lambda_1$ and $\lambda_2$ with
$\lambda_1<\lambda_2$, the difference between the corresponding phase
functions $\varphi_1$ and $\varphi_2$ will behave according to
\begin{equation}
  \varphi_2-\varphi_1 = \frac{(\lambda_2-\lambda_1)W_0}{5P_0^{\;2}(Y_0-3)^2}\,r^5 
  + O(r^7) > 0
\end{equation}
near the centre. Let $r_i>0$ be a radius sufficiently small for
$\varphi_1(r_i)<\varphi_2(r_i)$ to hold. This gives us the two first
order differential equations
\begin{align}
  \frac{d\varphi_1}{dr} &= F(r,\varphi_1,\lambda_1) \\
  \frac{d\varphi_2}{dr} &= F(r,\varphi_2,\lambda_2) \\
\end{align}
subject to the initial condition inequality
\begin{equation}
  \varphi_1(r_i) < \varphi_2(r_i). 
\end{equation}
Since
\begin{equation}
  F(r,\varphi_2,\lambda_2) \geq F(r,\varphi_2,\lambda_1)\quad \mbox{for $0<r<R$},
\end{equation}
it follows from a standard theorem on differential
inequalities\cite{andersson:ode} that
\begin{equation}
  \varphi_1 < \varphi_2 \quad \mbox{for $0<r<R$}, 
\end{equation}
given that $F(r,\varphi,\lambda)$ is Lipschitz continuous in $\varphi$
and continuous in $r$ on this interval. The former condition is
clearly satisfied since $F(r,\varphi,\lambda)$ has a simple
trigonometric dependence on $\varphi$. The possible complication that
$F(r,\varphi,\lambda)$ may have a finite number of jump
discontinuities in $r$ of the type
\begin{equation}
  F_- = \lim_{r\rarr r_b^-}F(r,\varphi,\lambda) \neq 
  F_+ = \lim_{r\rarr r_b^+}F(r,\varphi,\lambda), \qquad 
  \mbox{$|F_-|,\, |F_+| < \infty$, \: $0<r_b<R$}, 
\end{equation}
is not a problem. Indeed, since we assume that $\varphi$ is always to
be matched continuously across such radii $r_b$ it is very easy to
show that the theorem extends to a situation when $F(r,\phi,\lambda)$
has discontinuities in $r$ of this type. We have thus proved that
$\varphi(r,\lambda)$ is a strictly increasing function of $\lambda$
for all $r\in (0,R)$.

To prove the limit \refeq{lim1} we begin by claiming that it is not
difficult to convince oneself that given the expansions
\refeq{Pcenter} - \refeq{Qcenter} subject to the relation
\refeq{constrel}, there exist constants $r^\darr>0$ and
$\lambda^\darr>0$ for which the following statement is true: If
$P^\darr_0$ and $W^\darr_0$ are any constants satisfying
\begin{align}
  &0<P^{\darr-1}_0<P_0^{-1} \\
  &0<W^\darr_0<W_0
\end{align}
then for all $r\in I^\darr = (0,r^\darr]$, for all $\varphi$ and for
all $\lambda>\lambda^\darr$,
\begin{equation}\lbeq{Fdown}
  F(r,\varphi,\lambda) > P^{\darr-1}\cos^2\!\varphi + Y^\darr\sin{2\varphi} 
  + (Q^\darr+\lambda W^\darr)\sin^2\!\varphi =: F^\darr(r,\phi,\lambda),
\end{equation}
where
\begin{align}
  P^\darr &= P^\darr_0 r^{-2} \\
  Y^\darr &= Y_0 r^{-1} \\
  W^\darr &= W^\darr_0 r^{-2} \\
  Q^\darr &= Q^\darr_0 r^{-2}, \quad Q^\darr_0 = -P^\darr_0Y_0(3-Y_0).
\end{align}
This will be true if we can choose $r^\darr$ and $\lambda^\darr$ such
that, for all $r\in I^\darr$,
\begin{align}
  &P^{-1}-P^{\darr-1} > 0 \\ \lbeq{deltaQ}
  &Q-Q^\darr + \lambda(W-W^\darr) > 0 \\ \lbeq{detdelta}
  &(P^{-1}-P^{\darr-1})[Q-Q^\darr + \lambda(W-W^\darr)]-(Y-Y^\darr)^2 > 0.
\end{align}
To achieve this we first note that
\begin{align}
  P^{-1}-P^{\darr-1} &= (P_0^{-1}-P^{\darr-1}_0)r^2 + O(r^4) > 0 \\
  W-W^\darr &= (W_0-W^\darr_0)r^{-2} + O(r^0) > 0 
\end{align}
can be made valid on $I^\darr$ by choosing $r^\darr$ sufficiently
small. Furthermore, since the $r^{-4}$ coefficient of
\begin{equation}
  Q-Q^\darr = (P^\darr-P)Y_0(3-Y_0)r^{-4} + O(r^{-2}) 
\end{equation}
is positive (or zero if $Y_0=0$), we can choose $\lambda^\darr$
sufficiently large for the inequality \refeq{deltaQ} to hold. It now
follows directly from $Y-Y^\darr = O(r)$ that the remaining inequality
\refeq{detdelta} can also be made valid, since it shows that it
suffices to set $\lambda^\darr$ to an even larger number, if
necessary. Let us now consider the two differential equations
\begin{align}
  \frac{d\varphi}{dr} &= F(r,\varphi,\lambda) \\
  \frac{d\varphi^\darr}{dr} &= F^\darr(r,\varphi^\darr,\lambda),
\end{align}
and focus on the unique solutions that leave the centre according to
\begin{align}
   \varphi &= \frac1{P_0(3-Y_0)}\,r^3 + O(r^5) \\
  \varphi^\darr &= \frac1{P^\darr_0(3-Y_0)}\,r^3 + O(r^5). 
\end{align}
Since 
\begin{equation}
  \frac1{P_0(3-Y_0)} > \frac1{P^\darr_0(3-Y_0)},
\end{equation}
we can always, for every fixed value of $\lambda$, find $r_i\in
I^\darr$ such that
\begin{equation}
  \varphi(r_i,\lambda) > \varphi^\darr(r_i,\lambda). 
\end{equation}
It then follows from eq.\ \refeq{Fdown} and the already used theorem
on differential inequalities that
\begin{equation}
  \varphi(r,\lambda) > \varphi^\darr(r,\lambda), 
\end{equation}
for all $r\in I^\darr$ and all $\lambda>\lambda^\darr$. We now use
that $\varphi^\darr(r,\lambda)$ can be found in closed form and is
given by
\begin{equation}
  \tan\varphi^\darr = \frac1{P^\darr_0}
  \frac{\sin(kr)-kr\cos(kr)}{(kr)^2\sin(kr)-Y_0[\sin(kr)-kr\cos(kr)]}\,r^3,
  \quad k = \sqrt{\frac{W^\darr_0\lambda}{P^\darr_0}}. 
\end{equation}
Since we can hence infer that
\begin{equation}
  \lim_{\lambda\rarr\infty}\varphi^\darr(r,\lambda) = \infty
\end{equation} 
for every $r>0$, the same clearly is true for $\varphi(r,\lambda)$ for
every $r\in I^\darr$. Finally, we only need to note that since
$\varphi(r,\lambda)$ can never cross a line $\varphi = n\pi$
downwards, it follows that the result extends to the whole interval
$0<r<R$. 

Now, to prove the limit \refeq{lim2} we start off by instead bounding
$F(r,\phi,\lambda)$ from above near the centre, much in the same way
as when we bounded it from below by $F^\darr(r,\phi,\lambda)$. We thus
take two constants $P_0^\uarr$ and $W_0^\uarr$ such that
\begin{align}
  &P^{\uarr-1}_0 > P^{-1}_0 > 0 \\
  &W^\uarr_0 > W_0 > 0. 
\end{align}
Then there exist $r^\uarr>0$ and $\lambda^\uarr<0$ such that for all
$r\in I^\uarr = (0,u^\uarr]$, for all $\varphi$ and all
$\lambda<-\abs{\lambda^\uarr}$, it is true that
\begin{equation}\lbeq{Fup}
  F(r,\varphi,\lambda) < P^{\uarr-1}\cos^2\!\varphi + Y^\uarr\sin{2\varphi} 
  + (Q^\uarr +\lambda W^\uarr)\sin^2\!\varphi =: F^\uarr(r,\phi,\lambda),
\end{equation}
where
\begin{align}
  P^\uarr &= P^\uarr_0 r^{-2} \\
  Y^\uarr &= Y_0 r^{-1} \\
  W^\uarr &= W^\uarr_0 r^{-2} \\
  W^\uarr &= Q^\uarr_0 r^{-2}, \quad Q^\uarr_0 = -P^\uarr_0Y_0(3-Y_0).
\end{align}
Letting $\varphi(r,\lambda)$ and $\varphi^\uarr(r,\lambda)$ be the
unique solutions of
\begin{align}
  \frac{d\varphi}{dr} &= F(r,\phi,\lambda) \\
  \frac{d\varphi^\uarr}{dr} &= F^\uarr(r,\phi,\lambda),
\end{align}
that leave the centre as
\begin{align}
  \varphi(r,\lambda) &= \frac1{P_0(3-Y_0)}\,r^3 + O(r^5) \\
  \varphi^\uarr(r,\lambda) &= \frac1{P^\uarr_0(3-Y_0)}\,r^3 + O(r^5),
\end{align}
then, if $r\in I^\uarr$ and $\lambda<-\abs{\lambda^\uarr}$,
\begin{equation}
  \varphi(r,\lambda) < \varphi^\uarr(r,\lambda).
\end{equation}
For negative $\lambda$ the exact form for $\varphi^\uarr(r,\lambda)$
is given by
\begin{equation}
  \tan\varphi^\uarr(r,\lambda) = \frac1{P^\uarr_0}
  \frac{\sinh(kr)-kr\cosh(kr)}{(kr)^2\sinh(kr)+Y_0[\sinh(kr)-kr\cosh(kr)]}\,r^3,
  \quad k = \sqrt{\frac{W^\uarr_0\abs\lambda}{P^\uarr_0}},
\end{equation}
from which we can see that
\begin{equation}
  \lim_{\lambda\rarr-\infty}\varphi^\uarr(r,\lambda) = 0 \quad \mbox{for all $r>0$}. 
\end{equation}
The same is hence true for $\varphi(r,\lambda)$ for every $r\in
I^\uarr$ since it cannot cross the line $\varphi=0$ downwards.  Let us
now choose $\lambda$ sufficiently negatively large so that
$\phi(r^\uarr)<\epsilon$ for some $\epsilon\in(0,\pi)$. Let us also
assume that $\varphi(r_0,\lambda) = \epsilon$ for some $r_0\in
(r^\uarr,R)$. However, since $\sin\epsilon\neq 0$ we can make
$d\varphi/dr$ negative at $r=r_0$ by choosing $\abs\lambda$ even
larger, if necessary, implying that $\varphi(r,\lambda)$ cannot reach
the line $\varphi=\epsilon$ from below if $\lambda$ is sufficiently
negatively large. Thus we have proved the limit \refeq{lim2} since
$\epsilon$ can be chosen arbitrarily small.

\end{document}